# Mitigation of Cascading Outages Using a Dynamic Interaction Graph-Based Optimal Power Flow Model

CHANGSHENG CHEN [1], (Student Member, IEEE), WENYUN JU[2], (Member, IEEE), KAI SUN[3], (Senior Member, IEEE), AND SHIYING MA[1].

[1]China Electric Power Research Institute, Beijing 100192, China
[2]Electric Power Group, LLC, Pasadena, CA 91101, USA
[3]Department of Electrical Engineering and Computer Science, The University of Tennessee, Knoxville, TN 37996, USA

Corresponding author: Kai Sun (kaisun@utk.edu).

This work was supported in part by the National Natural Science Foundation of China (51777196) and in part by the ERC Program of the NSF and DOE under NSF grant EEC-1041877.

**ABSTRACT** Online decision support for effective mitigation actions against propagating cascading outages in a power grid still poses a big challenge to grid operators. This paper proposes an online mitigation strategy against cascading outages using an optimal power flow model based on a Dynamic Interaction Graph. For a given power grid, its interaction graph, also called an interaction model, was proposed in recent literature, which is composed of components and links that have large contributions to outage propagation. Differing from a conventional interaction graph, a Dynamic Interaction Graph can adaptively update its parameters with changes of grid topology and hence is more accurate in online identification of the most vulnerable transmission lines and likely outage propagation paths for mitigation. Further, the paper introduces an optimal power flow model based on the proposed Dynamic Interaction Graph for determining the optimal control strategy to maintain transfer margins of vulnerable transmission lines and effectively mitigate outage propagation. The numerical results on the IEEE 118-bus system demonstrate the effectiveness of the proposed models and associated control strategy.

## I. INTRODUCTION

Cascading outages are the leading causes of large-scale blackouts, e.g., the 2003 Northeast blackout [1], the 2011 Arizona-southern California blackout [2] and the 2012 Indian blackout [3]. In practice, online contingency analysis can reduce the risk of cascading outages by identifying potentially overloaded system components. However, it may not foresee all possible cascading outages since the operating condition keeps changing. Also, once cascading outages are initiated, network topology may dramatically change in an unpredictable manner [4], so online decision support to effectively mitigate propagating cascading outages still poses a big challenge to transmission system operators.

Because of the low-probability nature of cascading outages, limited historical data do not present a variety of ways in which cascading outages propagate in power systems. Alternatively, by using the cascading outage models, cascading outage data can be generated by computer simulations. These models are such as the hidden failure model [5], Manchester model [6], [7], CASCADE model [8], the collection of OPA models [9]-[15], dynamic and quasi-dynamic models [16]-[18], PRA model [19], sandpile model [20]. Analyzing these outage data using statistical tools provides a feasible way to understand the patterns in which outages propagate in a power grid. Ref. [21] analyzed the influence of weather on cascading outages. Ref. [22] and [23] studied how to fast identify vulnerable lines. Ref. [24]-[26] analyzed the impacts of cyber-attacks on power grids vulnerability to cascading outages, and Ref. [27], [28] studied mitigation and recovery approaches from cascading failures.

Recently, study cascading outages based on graph methods has been developed rapidly. In the early stage, topological network models in which the nodes denote buses and the links denote transmission lines are adopted to study cascading outages [29]-[31]. However, practical component outages can propagate non-locally and the next component to fail after a particular line outage can be distant. On the other hand, in order to model outage propagation paths, several graph-based models have been proposed, which can be constructed from simulated outage data, and can help analyze the ways how cascading outages propagate in a particular system. For instance, the influence graph used in [32] and [33] uses nodes of the graph to represent transmission lines, and edges to measure influences between outages of transmission lines. Ref. [34] proposes an interaction graph model to identify the critical components of a power system and key linkages of component failures under outage propagation using a dataset of cascading outages. This model is then improved in [35] by Expectation Maximization algorithm for more efficient computation. Ref. [36] future extends the single-layer interaction graph to a multi-layer interaction graph, where each layer focuses on one aspect of outage propagation, e.g., the number of line outages, the amount of load loss, and the electrical distance of the outage propagation. These interaction graph models provide an effective way to understanding the propagation mechanism of cascading outages, and are promising for online applications since the trend and consequences of outage propagation are immediately foreseen directly from an interaction graph without much computation efforts once it is constructed offline.

Thus, for effective online mitigation of ongoing cascading outages, an interaction graph model that has been built offline will need to be updated online to adapt to ongoing outages and take into account real-time operations. How to dynamically maintain a more adaptive interaction graph model and design associated mitigation strategy has not been studied in literature.

This paper extends the interaction graph model in [34] to a Dynamic Interaction Graph (DIG), which are constructed offline from historical or simulated cascading outage data and also updated online according to ongoing outages. The graph tells the key components of the power grid that play crucial roles in the propagation of outages.

In practice, there are plenty of approaches to alleviating transmission line overloads in grid operations, such as load shedding [37], generation re-dispatch [38], line switching [39], and flexible power flow adjustment of FACTS devices [40]-[43]. In particular, FACTS devices are very effective in mitigating transmission line overloads in both HVAC and HVDC networks. Any of the aforementioned approaches that reduce the tripping probabilities (overloading rate) of the dynamically changing key components are feasible to be applied in the proposed DIG-based online mitigation method. Since proactive generation re-dispatch is a promising and practical mitigation measure against cascading outages, this paper focuses on integration of generation re-dispatch into a mitigation strategy using the proposed DIG based optimal power flow (OPF) model.

The main contributions of this paper include:

1) The proposed DIG model is online updated with happening outages and changes in the operating condition and is more accurate in predicting propagation of outages;

2) The proposed DIG based OPF model includes a new inequality constraint on key components of the DIG, and can identify effective mitigation control actions for increasing transfer margins against a next outage.

The remainder of this paper is organized as follows. Section II briefly introduces the interaction graph model. Section III elaborates the proposed DIG model, the DIG based OPF model, and the associated mitigation strategy. Section IV demonstrates the proposed models and mitigation strategy on the IEEE 118-bus system using the OPA model. Section V discussed the feasibility of the proposed algorithms in real power systems. Finally, this work is concluded in Section VI.

## II. INTRODUCTION OF INTERACTION GRAPH

The interaction graph of a power system does not represent the actual topology of the grid. Its nodes represent the grid components, which are typically transmission lines and transformers, and its edges (or links) are used to measure influences between outages of components. There are the following three steps to build an interaction graph for a power system with $n$ components with risks of overloading under cascading outages [34]. The interaction graph builds upon a so-called interaction matrix $B$.

**Step 1** generates a database of cascading outage scenarios, which are called "cascades" in the rest of this paper. Any cascading outage simulation based on engineering principles can be used to produce the data needed to synthesize the interaction graph. Then group these data into different stages within each cascade based on the sequences or timing of outages. Assume that the size of the database is $K$ and $m$ represents the stage. Assume $M$ to be the maximum value of $m$. The grouped data can be illustrated as follows

|  | Stage 0 | Stage 1 | Stage 2 | . . . |
|---|---|---|---|---|
| Cascade 1 | $F_0^{(1)}$ | $F_1^{(1)}$ | $F_2^{(1)}$ | . . . |
| Cascade 2 | $F_0^{(2)}$ | $F_1^{(2)}$ | $F_2^{(2)}$ | . . . |
| $\vdots$ | $\vdots$ | $\vdots$ | $\vdots$ | $\vdots$ |
| Cascade $K$ | $F_0^{(K)}$ | $F_1^{(K)}$ | $F_2^{(K)}$ | . . . |

**Step 2** constructs the interaction matrix $B$. Firstly, construct a matrix $A \in \mathbb{Z}^{n \times n}$ whose entry $a_{ij}$ is the number of times that component $i$ fails in one stage before the failure of component $j$ among all cascades. $A$ cannot be used as the interaction matrix directly since it exaggerates the interactions between component failures, i.e., it asserts one component interacts with another one only because it fails in its last stage. Therefore, the causal relationships between failed components should be determined. Specifically, for any two consecutive stages, $m$ and $m+1$ of any cascade $x$, assume that the set of components in stage $m$ is $C_m$, and component $j$ failed in stage $m+1$. Then the component whose failure is considered to cause the failure of component $j$ can be determined through Eq. (1).

$$I_c = \left\{ i_c \mid i_c \in C_m \text{ and } a_{i_c j} = \max_{i \in C_m} a_{ij} \right\} \quad (1)$$

Fig. 1 illustrates the determination process of 2 consecutive stages of a cascade. Assume that the values of the edges satisfy Eq. (2) and Eq. (3).

$$a_{AD} = a_{BD} = \max_{i \in \{A,B,C\}} a_{iD} \quad (2)$$

$$a_{CE} = \max_{i \in \{A,B,C\}} a_{iE} \quad (3)$$

From Fig. 1(a) to Fig. 1(b), the edges $A \rightarrow E$, $B \rightarrow E$, and $C \rightarrow D$ are removed, i.e., the elements $a_{AE}$, $a_{BE}$, and $a_{BE}$ are corrected to be 0. Therefore, the causal relationships are the failures of $A$ and $B$ cause the failure of $D$, and the failure of $C$ causes the failure of $E$.

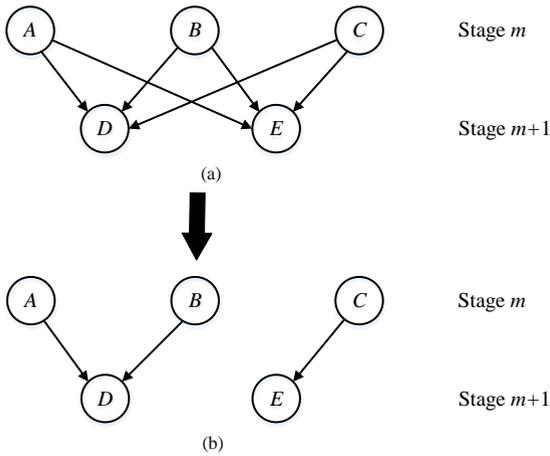

**FIGURE 1.** Determining process of two consecutive stages of a cascade.

After determining the causal relationships for all cascades, $A$ can be corrected to be $A' \in \mathbb{Z}^{n \times n}$, whose entry $a'_{ij}$ is the number of times that the failure of component $i$ causes the failure of component $j$. Then the interaction matrix $B \in \mathbb{Z}^{n \times n}$ can be obtained from $A'$. Its entry $b_{ij}$ is the empirical probability that the failure of component $i$ causes the failure of component $j$, which is given by

$$b_{ij} = \frac{a'_{ij}}{N_i} \quad (4)$$

where $N_i$ is the number of failures of component $i$.

*Step 3* builds the interaction graph model. The interaction matrix $B \in \mathbb{Z}^{n \times n}$ determines how components interact with each other. The nonzero elements of $B$ are called links. For instance, link $l : i \rightarrow j$ corresponds to $B$'s nonzero element $b_{ij}$ and starts from component $i$ and ends with component $j$. By putting all links together, an interaction graph denoted by $G(C, L)$ can be obtained. Its vertices $C$ are components, and each directed link $l \in L$ represents that a failure of the source vertex component causes the failure of the destination vertex component with a probability $b_{ij}$.

## III. PROPOSED DIG BASED MITIGATION STRATEGY

### A. PROPOSED DYNAMIC INTERACTION GRAPH

In this paper, the motivation for improving an interaction graph model to a DIG model is to make the graph be adaptively updated to reflect ongoing outages and online data so that the trend and consequences of outages can be more accurately predicted. That is vitally important for determining effective mitigation actions. Unlike existing interaction graph models, a DIG can update its topology, key components, key linkages and other parameters along with the propagation of outages.

Define $B'_m$ as the dynamic interaction matrix on stage $m$. Assume that there is no component fails in stage 0, i.e., $B'_0 = B \in \mathbb{Z}^{n \times n}$, and the initial faults happen starting from stage 1. With the propagation of cascading outages, $B'_m$ ($m > 0$) needs to be corrected simultaneously. Assume that totally $q$ components fail after $m$ stages, whose indices form set $Q$. Thus, $B'_m$ can be obtained by removing the rows and columns corresponding to $Q$ from $B'_0$. Each row indicates the influence of the failure of one component on the others and each column indicates the influences of the failures of the other components on this component. The dimension of $B'_m$ is reduced to $(n - q) \times (n - q)$. Its element $b'_{ij}$ is given by

$$b'_{ij} = \left\{ b_{ij} \mid i \notin Q, j \notin Q \right\} \quad (5)$$

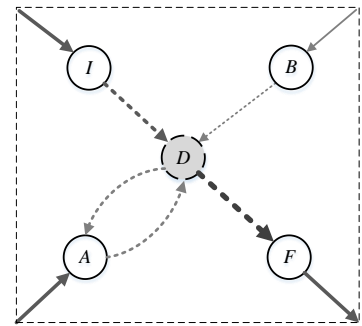

**FIGURE 2.** A part of the DIG.

Fig. 2 illustrates a part of the DIG, and the thickness of the links represents the link weights. Assume that component $D$ fails in stage $m$-1, and then the dynamic interaction matrix $B'_m$ can be obtained by removing all the links pointing to $D$, i.e., $I \rightarrow D$, $B \rightarrow D$, and $A \rightarrow D$, and all the links start from $D$, i.e., $D \rightarrow A$ and $D \rightarrow F$.

The mathematical formulation of dynamic interaction matrix $\boldsymbol{B}'_m$ on the condition of $D$ fails is given by

$$\boldsymbol{B}'_m = \begin{bmatrix} b_{11} & \cdots & b_{1(D-1)} & b_{1(D+1)} & \cdots & b_{1n} \\ \vdots & \cdots & \vdots & \vdots & \cdots & \vdots \\ b_{(D-1)1} & \cdots & b_{(D-1)(D-1)} & b_{(D-1)(D-1)} & \cdots & b_{(D-1)n} \\ b_{(D+1)1} & \cdots & b_{(D+1)(D-1)} & b_{(D+1)(D+1)} & \cdots & b_{(D+1)n} \\ \vdots & \cdots & \vdots & \vdots & \cdots & \vdots \\ b_{n1} & \cdots & b_{n(D-1)} & b_{n(D+1)} & \cdots & b_{nn} \end{bmatrix} \quad (6)$$

Finally, a DIG $\boldsymbol{G}'(\boldsymbol{C}', \boldsymbol{L}', m)$ can be obtained based on $\boldsymbol{B}'_m$. Note that the DIG downgrades to a static interaction graph when $m = 0$, i.e., $\boldsymbol{G}'(\boldsymbol{C}', \boldsymbol{L}', 0) = \boldsymbol{G}(\boldsymbol{C}, \boldsymbol{L})$.

### B. THE QUANTIZATION OF DIG

After deriving a DIG, how to online use its information for mitigation of cascading outages is the next problem. Considering generation re-dispatch as a feasible mitigation strategy, we may utilize quantitative information carried by DIG in an OPF model for solving the control strategy. This section will focus on how to obtain useful quantitative information from a DIG.

For each stage $m$, define index $c_{i,m}$ as the expected number of outages that caused by the failure of component $i$. $c_{i,m}$ can be calculated through a unique directed acyclic subgraph which can be extracted from the DIG. Fig. 3 illustrates the process of obtaining the directed acyclic subgraph $\boldsymbol{G}'(i, m)$ regarding each component. Assume that there are more than 2 stages in this cascade, i.e., $M \geqslant 2$ (the subgraphs of $m > 1$ are not shown here).

Fig. 3(a) is a subgraph extracted from DIG when $m = 0$. From Fig. 3(a) to Fig. 3(b), vertices $H$, $I$, and $J$, as shown by dashed circles, are removed since they have no paths from $i$ or in other words are not influenced by the failure of $i$. Then these two types of edges (indicate by dotted arrows) are removed: ones coming from vertices of a future stage such as $A \rightarrow D$ and $E \rightarrow i$, for causality between different stages, and the ones coming from vertices of the same stage such as $B \rightarrow D$, for independency in the same stage. Then the directed acyclic subgraph $\boldsymbol{G}'(i, 0)$, i.e., Fig. 3(b), is obtained for which there is no loop and for each vertex, there is exactly one vertex (i.e., the cause) pointing to it.

Assume that component $D$ fails in stage 1. From Fig. 3(b) to Fig. 3(c), i.e., $\boldsymbol{G}'(i, 1)$, the vertices which $D$ pointing to ($A$, $F$, dashed circles) and the corresponding links ($J \rightarrow D$, $D \rightarrow A$, $D \rightarrow F$, dotted arrows) are removed since $D$ cannot influence other components or be influenced. Therefore, $\boldsymbol{G}'(i, m)$ will be dynamic varying with the operational condition when $m \geqslant 1$.

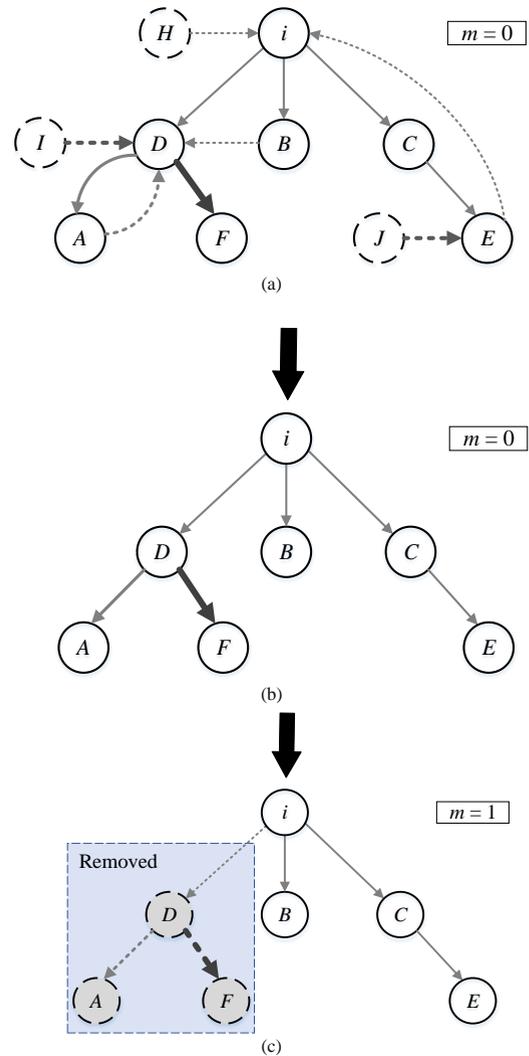

**FIGURE 3.** Diagram for obtaining the directed acyclic subgraph starting from $i$.

Based on the directed acyclic subgraph $\boldsymbol{G}'(i, m)$, the index $c_{i,m}$ for each component can be calculated by

$$c_{i,m} = \sum_{k \in \boldsymbol{C}'(i,m)} E_s b_{sk} \quad (7)$$

where $\boldsymbol{C}'(i, m)$ is the set of vertices in $\boldsymbol{G}'(i, m)$. $b_{sk}$ is the $s^{\text{th}}$ row and $k^{\text{th}}$ column element of matrix $\boldsymbol{B}'_m$. $E_s$ is the expected number of outages of the source vertices that pointing to vertex $k$. Therefore, $E_s b_{sk}$ gives the expected value of outages propagates from link $s \rightarrow k$. Note that, since no source vertices are pointing to vertex $i$, $E_i$ is set to be $N_i / N_0$, i.e., the empirical probability of the failure of component $i$, where $N_0$ is the size of the database.

By taking $c_{i,m}$ as the weights of components, DIG can be transformed into a directed weighted graph. The greater $c_{i,m}$ is, the more critical the component is for the propagation of cascading outages. The top-ranked components are defined as key components $C^{key}$. Key components $C^{key}$ contains the critical information of DIG.

### C. PROPOSED DIG-OPF MODEL

Eq. (8) to (13) are the mathematical formulation of a classical DC-OPF model, which is widely used in literature for simulation and control design with cascading outages, e.g., the OPA model and its many variants. In the rest of the paper, this DC-OPF model is referred to as a classical OPF model for comparison purposes.

$$\min \sum_{i \in G} c_i p_i + \sum_{i \in L} W_i |p_i - P_i| \qquad (8)$$

S.t.

$$\boldsymbol{F} = \boldsymbol{A}\boldsymbol{p} \qquad (9)$$

$$\sum_{i=1}^{n} p_i = 0 \qquad (10)$$

$$P_i \leq p_i \leq 0, i \in L \qquad (11)$$

$$0 \leq p_i \leq P_i^{\max}, i \in G \qquad (12)$$

$$-\boldsymbol{F}_{\max} \leq \boldsymbol{F} \leq \boldsymbol{F}_{\max} \qquad (13)$$

where $G$ and $L$ denote the generator set and load set respectively. $c_i$ is the generation cost coefficient for generator $i$. The coefficient $W_i$ is the economic loss for the load $i$. $p_i$ is the active power injected at bus $i$, and $\boldsymbol{p} = (p_1, ..., p_n)^T$ is the vector of active power injections. $P_i$ is the demand power at bus $i$, and $P_i^{max}$ is the maximum power limit for generator $i$. $\boldsymbol{F}$ and $\boldsymbol{F}_{max}$ are the vectors of line flows and their limits.

Define $M_{ij} = F_{ij} / F_{ij,max}$ as the overloading rate of the line connecting the nodes $i$ and $j$, where $F_{ij}$ and $F_{ij,max}$ are the power flow and line flow limit of the line. A line with $M_{ij} < 1$ still has a margin to carry more power, but it may trip due to the unwanted operation of relay protection. The higher the $M_{ij}$ is, the larger the probability of unwanted operation is. On the other hand, a line with $M_{ij} \geq 1$ has no margin and has a probability of being tripped due to violating or bounding its thermal limit.

The objective function, Eq. (8), is aiming to minimize the cost, and no constraints in the classical OPF model take into account the differences of vulnerabilities of transmission lines. Thus, the solution of the classical OPF model cannot take into account the consequences of the outages of components. Sometimes, the pursuit of a better economy may drive high $M_{ij}$ of one or more vulnerable transmission lines and even increase the risk of cascading outages.

In order to improve the inadequacy of the classical OPF model, this paper proposed a novel DIG based OPF model (for short, DIG-OPF model) which can effectively reduce the risk of cascading outages. The mathematical formulation of the DIG-OPF model is the combination of the classical OPF model and the Eq. (14) given below. Eq. (14) is a new inequality constraint to take into account the consequences of the outages of key components.

$$-\alpha \boldsymbol{F}_{\max}^{key} \leq \boldsymbol{F}^{key} \leq \alpha \boldsymbol{F}_{\max}^{key} \qquad (14)$$

where $\boldsymbol{F}^{key}$ and $\boldsymbol{F}_{\max}^{key}$ are the vectors of line flows and the limits of key components. $\alpha$ is a scaling coefficient of 0 to 1.

With constraint (14), the power flows of the key components (i.e., transmission lines) are required to have a certain margin depending on $\alpha$ to the limits. In this way, the tripping probabilities of the key components will be greatly reduced. Since these key components play crucial roles in the propagation of cascading outages, reduce their tripping probabilities can significantly reduce the risk of large-scale cascading events. Therefore, the proposed DIG-OPF model is far more efficient than the classical OPF model in mitigating cascading outages. Note that, because key components are identified based on the DIG and will be updated with the change of network topology for accuracy, constraint (14) is in fact applied to a varying set of lines.

An interpretation of constraint (14) is emulation of this practical operation: when cascading outages occur, the operator pays special attentions to a few more vulnerable lines and intentionally limit their power flows to reduce the overloading rate as well as the probability of unwanted tripping. Such a set of lines dynamically change at different stages of outages as told by the key components of the DIG.

Note that a future power system will become more flexible through continuous penetration of power electronic devices such as FACTS and HVDC systems. These devices should be considered and modeled. Although the present version of the DIG-OPF model has not yet considered these devices, in theory, they can be added to the algorithm, which requires adding new equations and constraints to the problem. Some existing studies, e.g., [40]-[43], give useful guidance on how to model and add these devices.

### D. PROPOSED DIG BASED MITIGATION STRATEGY

Fig. 4 presents the flowchart of the proposed DIG based mitigation strategy, and it performs the following steps.

**Step 1** generates the initial matrix $\boldsymbol{B}'_0$ offline based on a database of cascading events.

**Step 2** detects the current network topology based on online data. The data can be provided by the Supervisory Control and Data Acquisition (SCADA) system or algorithms based on real-time wide-area measurements.

**Step 3** updates the interaction matrix $\boldsymbol{B}'_m$ using (5) and (6) to reflect the current topology.

**Step 4** ranks the components by the index $c_{i,m}$ which can be calculated by (7), and then picks up a certain number, e.g., 20, of the top-ranked components as the key components ($C^{key}$).

**Step 5** solves the proposed DIG-OPF, i.e., Eq. (8)-(14).

**Step 6** generates a list of re-dispatch generators. Compare the presently scheduled generator outputs (obtained in the previous iteration of the algorithm) with the solution in *Step 5*, and the generators whose output changes, form the re-dispatch list. The corrective outputs of these re-dispatch generators are also based on the solution derived in *Step 5*.

*Step 7* performs mitigation control to increase the transfer margins of vulnerable lines. According to the solutions of DIG-OPF, the operators re-dispatch the generators provided by *Step 6*.

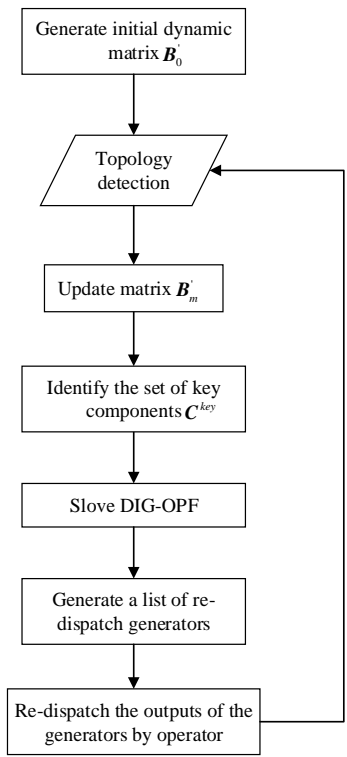

**FIGURE 4.** The flowchart of the proposed mitigation strategy.

Note that the optimal solution of the OPF algorithm usually requires to adjust the output levels of a large number of generators. This property is particularly undesirable in the current industry practices since the system operators may only be able to handle a limited number of corrective actions due to time and communication constraints. However, this issue itself, i.e., how to acquire a proper list of re-dispatch generators, is an important research topic in the study of security-constrained OPF, and a series of papers have discussed and proposed methods to resolve it [44]-[46]. Therefore, this paper focuses on studying the application of DIG in mitigating cascading outages and assume the operators are capable of handling all corrective actions provided by *Step 6*.

## IV. CASE STUDY ON THE OPA MODEL

The proposed DIG model and associated mitigation strategy are tested on the IEEE 118-bus test system using the OPA model in MATLAB 2018b environment. The OPA model contains two layers of iterations. The inner iterations are concerned with power-flow based simulation of grid operations under outages, referred to as "fast dynamics", and the outer iterations are about the long-term planning process, referred to as "slow dynamics", which simulate the growth and upgrading of the transmission network with the increases of generation and load. In this paper, only the inner iterations are considered.

In the rest of the paper, the OPA model and its settings are based on [13]. The parameters are set as follows: The initial failure probability of each line is assumed to be 0.01. The probability of tripping overloaded lines is set to be 0.999. The probability of tripping normal lines is set to be $0.001 \times (M_{ij})^n$, where 0.001 is the base probability of unwanted operation of relay protection. $n$ is set to be 10 since the higher the load rate is, the larger the probability of unwanted operation is. The coefficient $\alpha$ in the proposed DIG-OPF model is set to be 0.85, and the explanation of how to set $\alpha$ will be given in Section V.C.

### A. GENERATE THE DATABASE AND BUILDING IG (DIG FOR M=0)

Generate a database with 10,000 cascades using the OPA model. The conventional interaction graph (for short, IG, i.e., the DIG with $m$=0) can be derived base on the three steps in Section II.

Rank the components by index $c_{i,m}$. Pick up the top 20 components as the key components which are shown in Tab. 1 and Fig. 5(a). Line 92-89 in Tab. 1 has two parallel lines.

TABLE I
KEY COMPONENTS OF IG (DIG WITH M=0)

| Line | Line | Line | Line |
| --- | --- | --- | --- |
| 66-62 | 77-75 | 32-31 | 17-15 |
| 92-89 (1) | 92-89 (2) | 19-18 | 89-85 |
| 72-24 | 100-99 | 70-24 | 70-69 |
| 34-19 | 42-40 | 103-100 | 42-41 |
| 105-104 | 92-91 | 37-33 | 66-49 |

### B. DYNAMIC UPDATES OF KEY COMPONENTS WITH A SPECIFIC CASCADE

For a specific cascade, the key components of the DIG should vary with the changing of network topology. Fig. 5 illustrates the dynamic changing process of a specific cascade. Tab. 2 gives the failed components.

In stage 0, there was no component failed. The DIG has the same key components as the IG. In stage 1, there were 6 components outages, which caused a significant change of the key components. From Fig. 5(a) to (b), there are 4 key components changed. From Fig. 5(b) to (c), the number increases to 9 dues to the 6 outages in stage 2.

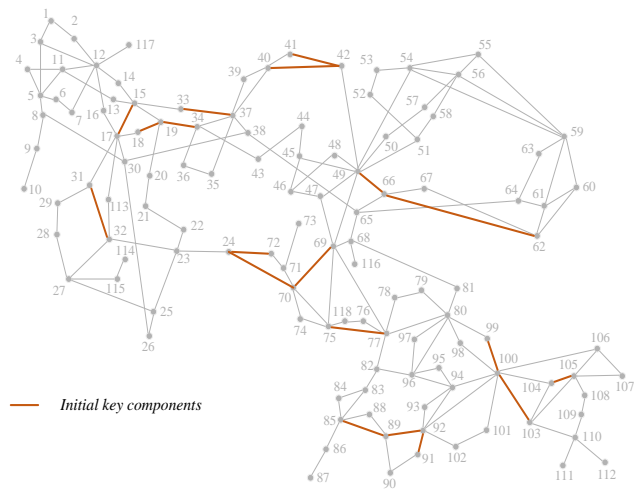

(a) Key components in stage 0.

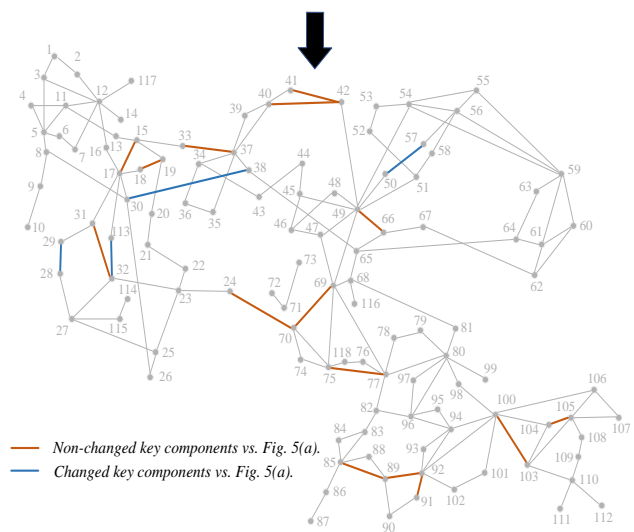

(b) Updated key components in stage 1.

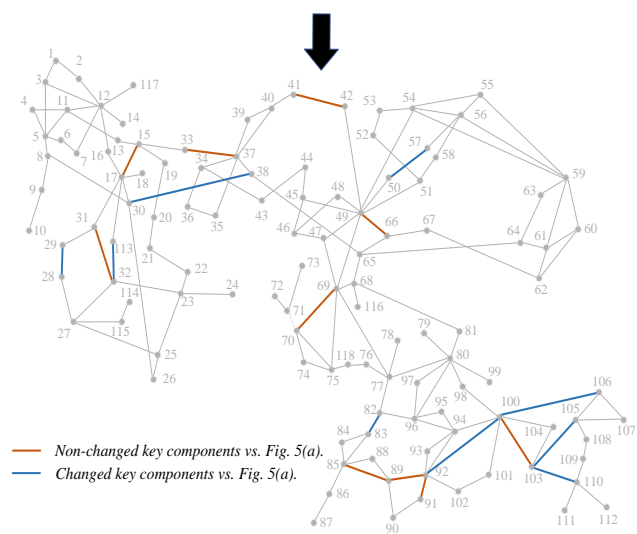

(c) Updated key components in stage 2.

**FIGURE 5.** The variation of key components of a specific cascade.

TABLE II
FAILED COMPONENTS

| m | Line | Line | Line | Line | Line | Line |
|---|------|------|------|------|------|------|
| 0 | | | NONE | | | |
| 1 | 15-14 | 34-19 | 66-62 | 71-70 | 72-24 | 100-99 |
| 2 | 19-18 | 42-40 | 70-24 | 77-75 | 79-78 | 105-104 |

Since the key components are essential to the proposed DIG-based mitigation strategy, the average number of the change of key components in each stage are calculated and shown in Fig. 6. The results indicate that the key components can change significantly, which cannot be easily neglected. Moreover, the average number increases as the stage increases. It is reasonable since the change of key components is caused by the change of network topology. The more stages usually lead to more outages.

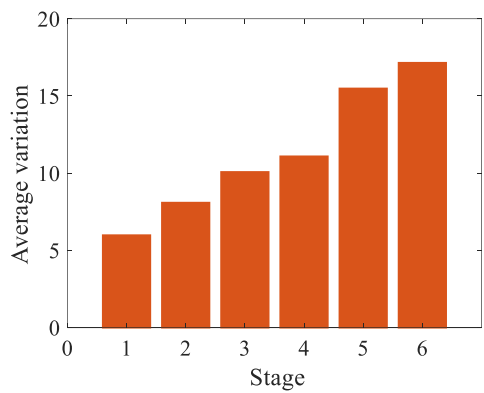

**FIGURE 6.** The average number of changed key components of each stage.

### C. PERFORMANCE OF THE PROPOSED MITIGATION STRATEGY

The performance of the proposed DIG based mitigation strategy is tested and compared with other three mitigation strategies in this section. All the cases use the OPA model to simulate, but the mitigation modules are different. The details are as follows:

*Case A*: The mitigation module uses the classical OPF model, i.e., Eq. (8)-(13), which is remarked as *classical-OPF*.

*Case B*: The mitigation module uses the DIG-OPF model, i.e., Equation (8)-(14), but the key components in constraint (14) are randomly selected. This case is remarked as *random-OPF*.

*Case C*: The mitigation module uses the DIG-OPF model, but the key components in constraint (14) are selected based on IG (the DIG with $m=0$), which is remarked as *IG-OPF*.

*Case D*: The mitigation module uses the DIG-OPF model, and the key components in constraint (14) are selected based on DIG, which is remarked as *DIG-OPF*.

Fig. 7 gives the Probability Distribution Function (PDF) curves of the four cases about line outages. The procedure of calculating PDF is as follows: First, set the bin width of continuous variable *x*. Then, count the number of cascades that variable *y* (*y* represents the number of lines failed) falls into one bin.

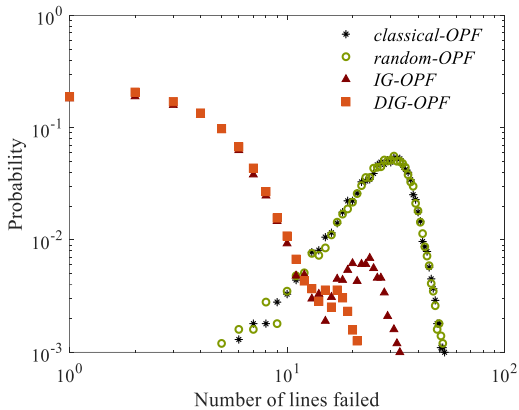

**FIGURE 7.** The PDF of line outages. (*classical-OPF* uses the classical OPF model, i.e., Eq. (8)-(13); *random-OPF* uses the DIG-OPF model, i.e., Equation (8)-(14), but the key components in constraint (14) are randomly selected; *IG-OPF* uses the DIG-OPF model, but the key components in constraint (14) are selected based on IG; *DIG-OPF* uses the DIG-OPF model, and the key components in constraint (14) are selected based on DIG)

In Fig. 7, the curve of *random-OPF* coincides with the curve of *classical-OPF*, which means *random-OPF* is no better than *classical-OPF*. It is reasonable since most of the randomly selected components are not crucial in the propagation of cascading events, thus reduce the failure probabilities of them cannot enhance the ability to mitigate cascading outages. However, when analysis the result of *IG-OPF*, it is obvious that the probability of small-scale cascades in *IG-OPF* is much higher than those in *classical-OPF*, whereas the probability of the large-scale cascades in *IG-OPF* is much lower than those in *classical-OPF*. Therefore, compare to *classical-OPF*, *IG-OPF* can significantly enhance the ability against cascading outages. Finally, as we expected, the probability of large-scale cascades in *DIG-OPF* is even lower than those in *IG-OPF*, which means *DIG-OPF* is more efficient than *IG-OPF*, especially for preventing large-scale cascades.

Plot the number of large-scale cascades which has more than 10 outages in Fig. 8. All the results are normalized to Case A. Fig. 8 indicates that the value of *random-OPF* is almost the same with *classical-OPF*, whereas *IG-OPF* reduces 90.48% large-scale cascades. Finally, the best performance one is *DIG-OPF* since it reduces 96.5% of large-scale cascades.

Since the amount of load shedding is another important metric of cascading events, the statistical results of load loss are also given. The average load shedding of each cascade is shown in Fig. 9. The number of cascades, which lost more than 5% of the total load demand, is given in Fig. 10. All the results are normalized to Case A.

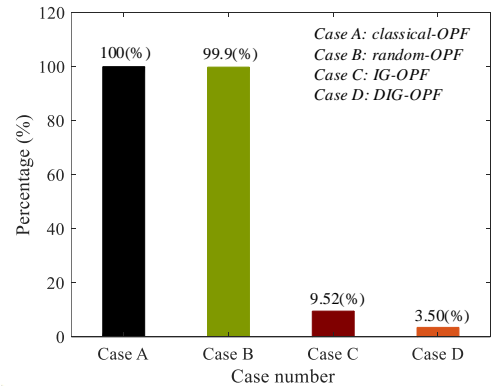

**FIGURE 8.** The number of large-scale cascades. (more than 10 outages)

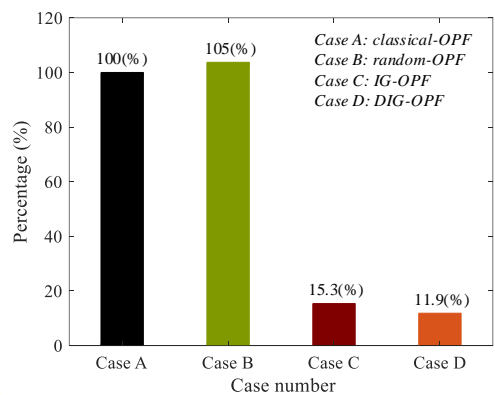

**FIGURE 9.** The average load shedding of each cascade.

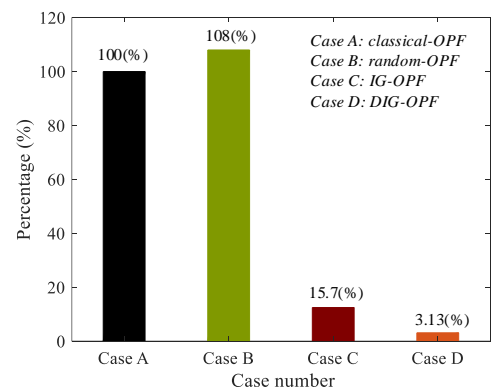

**FIGURE 10.** The number of cascades, which lost more than 5% of the total load demand.

The results in Fig. 9 and Fig. 10 are consistent with the results of the outages, i.e., both *IG-OPF* and *DIG-OPF* can significantly mitigate the load loss, but *DIG-OPF* is more efficient than *IG-OPF*.

The excellent performance of *DIG-OPF* attributed to the addition of constraint (14) and the well selection of key components. The key components selected based on DIG are the most crucial ones whose outages may cause serious consequences. By using constraint (14) to reduce the failure probabilities of the key components, the risky of large-scale cascades can be greatly reduced. Moreover, *DIG-OPF* has better performance than *IG-OPF* since the DIG more accurately reflect the changing vulnerabilities of the power network than IG.

## V. DISCUSSION ON THE FEASIBILITY OF USING DIG-OPF IN REAL POWER SYSTEMS

### A. COMPUTATION TIME AND CONVERGENCE OF DIG-OPF

Despite many excellent studies, e.g., [47] and [48], solving OPF problems with AC power flow constraints is still a major challenge in power system analysis due to convergence and time-consuming issues. At the same time, OPF with DC power flows is widely used in grid operations by the power industry. The above considerations motivate us to use the DC power flow model in the OPF problem in this paper. Therefore, the proposed DIG-OPF algorithm has the merit of DC OPF, i.e., it has more robust convergence and can be solved very fast. For *Case D* in Section IV, the proposed algorithm is tested on a PC configured with Intel (R) Core i7-6700 CPU, 3.4 GHz, and 16 GB RAM, and the average computation time is within 2 seconds, which is acceptable in practical applications.

### B. TESTS ON AN AC POWER FLOW MODEL

In Section IV, we have used the DC-OPA model to test the proposed mitigation strategy and explain its mechanism. However, since the DC-OPA model is based only on DC power flow, the feasibility of the proposed method in the real systems should be further discussed. To address this concern, a new AC power flow model is built, and its flowchart is shown in Fig. 11. This new model is referred to as the "AC test model" in the rest of the paper.

In Fig. 11, only the proposed control module still uses DC power flow (in the blue box), all other parts use AC power flow to more accurately simulate the real power system. By using the AC test model, the response of the systems can be well simulated after the operator re-dispatched the generators based on DIG-OPF. Therefore, the effectiveness of the proposed mitigation strategy considering AC power flows is tested.

Because this paper focuses on studies of cascading outages mainly caused by component failures due to overloading, it should be noted that both the DC and AC power flow models are steady state models ignoring transient dynamics. Moreover, a simplified load shedding scheme is added to the AC test model to coordinate with the proposed mitigation strategy. More practical load shedding schemes, such as the schemes in Ref. [49] and Ref. [50], can easily be integrated into the AC test model.

Test the proposed mitigation strategy on IEEE 118 system using the AC test model. The load factor is set to be 1.0. The parameters of DIG-OPF are set as follows: the scaling factor α in (14) is set to be 0.7, the explanation of how to choose α will be given in Section V.C. Other parameters are the same as the setting in Section IV.

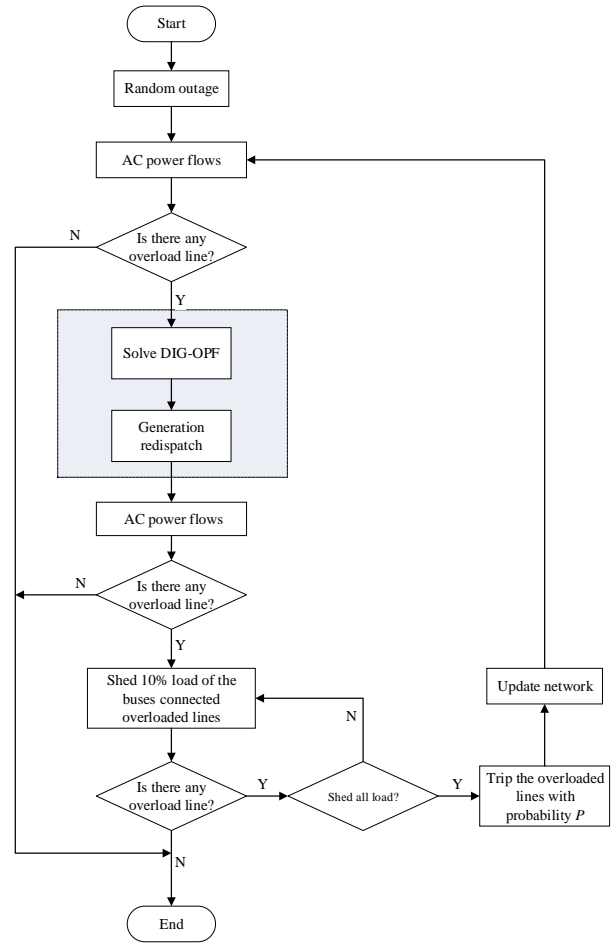

**FIGURE 11.** The flowchart of the AC test model.

Generate a database with 10,000 cascades using the AC test model, then change the DIG-OPF to classical-OPF and regenerate 10,000 cascades to compare results. All the statistical results are normalized to the classical-OPF. The comparison results are given in Fig. 12.

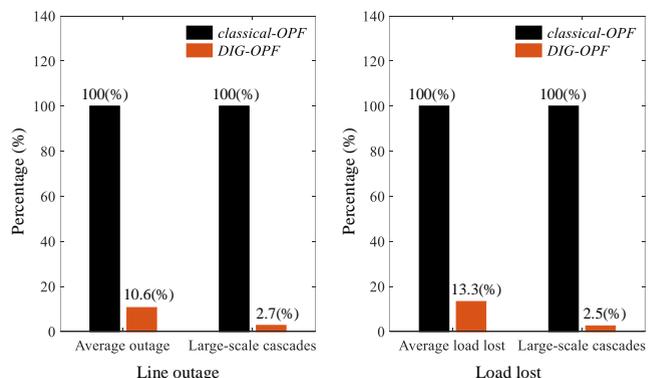

**FIGURE 12.** The comparison results of line outage and load lost, respectively.

Fig. 12 shows that the proposed DIG-OPF based mitigation strategy reduces 89.4% average line outages, 97.3% large-scale cascades (line outage), 86.7% average load lost, and 97.5% large-scale cascades (load lost). These results prove that the proposed method has potential in applications for real power systems.

### C. DISCUSSION OF SETTING THE SCALING FACTOR α

The scaling factor α was set to be 0.85 and 0.7 in Section IV and Section V.B, respectively, but the explanation of how to set this coefficient has not been given yet.

According to the analysis in Section III.C, the mechanism of mitigating cascading outages by using DIG-OPF is that it reduces the tripping probabilities of key components through constraint (14). Since DIG-OPF based on DC power flow, its calculating results may not very accurate. Therefore, the scaling factor α should be lower enough to ensure the power flows on the vulnerable lines within their limits.

In the range of 0 to 1, the value of α is set to be 0.05, 0.10, 0.15, ···, and 1.00, at intervals of 0.05. Then using the AC test model to generate 20 databases, and each database corresponds to a different α. The statistical results are shown in Fig. 13 and Fig.14. All the results are normalized to the classical-OPF.

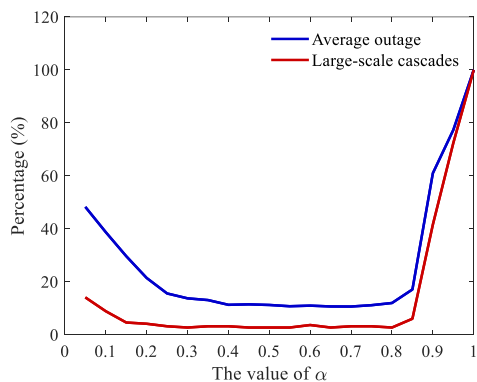

**FIGURE 13.** The comparison results of line outage when α changes.

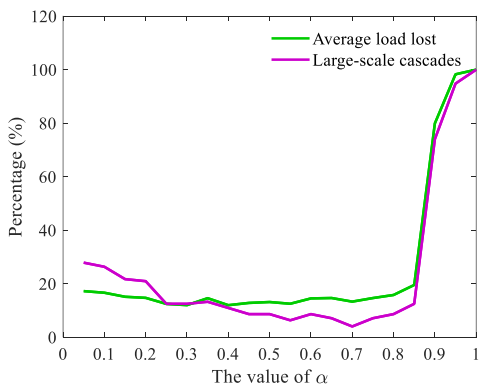

**FIGURE 14.** The comparison results of load lost when α changes.

In Fig. 13 and Fig. 14, all the curves cliffy drop when α decreases from 1.00 to 0.85. It is reasonable since the difference between the results of DC power flow and AC power flow is usually not very large, and setting α below 0.85 can effectively limit most of the line flows within their limits. On the other hand, the curves increase with the decrease of α when α is smaller than 0.4. That is because when the DIG-OPF algorithm intentional increases the transfer margins of the vulnerable lines, it also drives the power flows on other lines more likely to violate their limits. Therefore, setting α between 0.4 and 0.85 can make the proposed mitigation strategy perform best.

### D. DISCUSSION OF THE IMPACTS OF RENEWABLE ENERGY SOURCES

If a future power grid contains a majority of intermittent, undispatchable renewable energy sources, that may affect the accuracy of the proposed algorithm since the algorithm assumes a power system based mainly on conventional power plants, which can be dispatched by utility companies or system operators. Variable outputs of renewable energy sources will bring uncertainties to success of the proposed mitigation strategy. However, this is a challenging problem faced by many online applications for power grid operations and control in both normal and abnormal conditions. It is envisioned that uncertainties with renewable energy resources could be reduced or compensated by increasing energy storage devices or by new control techniques allowing renewable generations to operate in an inertial emulation mode like synchronous machines. Also, the power electronic interfaces allow more flexible control of both active and reactive powers to help support the grid. Thus, they can become more dispatchable and controllable under emergency conditions to cascading outages.

## VI. CONCLUSION

This paper proposed a DIG model to make the interaction graph model more adaptive to ongoing outages such that outage propagation can be mitigated online. The parameters of DIG can vary with the changing of network topology, which makes key components of the DIG more accurately reflect the changing vulnerabilities of the power network.

Based on DIG, a novel mitigation strategy was proposed. In this strategy, the DIG is defined as a constraint and then added to the classical OPF model for generation re-dispatch to increase transfer margins of vulnerable lines. The numerical results on the IEEE 118-bus system have shown that the DIG based control strategy can effectively mitigate cascading outages.

In practice, the proposed methods can give useful guidance to the operators. First, the DIG model can dynamically identify the vulnerable components online, thus the operators can save their attention to a small set of components. Second, the proposed DIG-OPF model can provide reference solutions that the operators can refer to for reducing the risk of cascading outages.

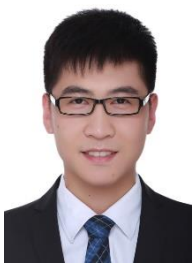

**CHANGSHENG CHEN** (S'18) received the B.S. degree from Yanshan University, Hebei, China, in 2013, and the M.S. degree in electrical engineering from North China Electric Power University, Hebei, China, in 2016. He is currently pursuing the Ph.D. degree in China Electric Power Research Institute, Beijing, China.

His research interests include cascading failure analysis and transient stability assessment.

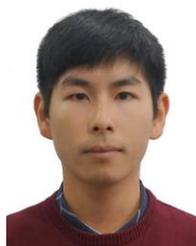

**WENYUN JU** (S'15-M'18) received the B.E. degree in electrical information from Sichuan University, Chengdu, China in 2010, M.Sc. degree in electrical and electronic engineering from Huazhong University of Science and Technology, Wuhan, China in 2013, and Ph.D. degree at the Department of EECS, University of Tennessee, Knoxville, TN, USA in 2018. He is now working in Electric Power Group, LLC as a Power System Engineer.

He is mainly involved in research and development related to real-time wide area monitoring with synchrophasor data, cascading outage, and generator model validation.

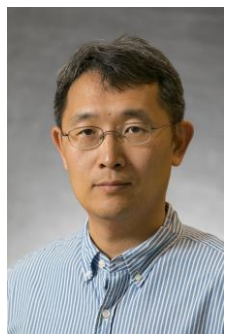

**KAI SUN** (M'06-SM'13) received the B.S. degree in automation and the Ph.D. degree in control science and engineering from Tsinghua University, Beijing, China, in 1999 and 2004, respectively. He was a Project Manager in grid operations and planning with EPRI, Palo Alto, CA, USA, from 2007 to 2012. He is currently an Associate Professor with the Department of EECS, the University of Tennessee, Knoxville, TN, USA.

His research interests include stability, dynamics and control of power grids, and other complex systems. Prof. Sun serves in the editorial boards of IEEE Transactions on Power Systems, IEEE Transactions on Smart Grid and IEEE Access.

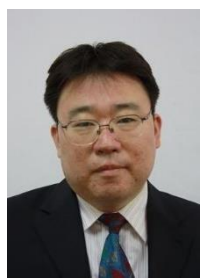

**SHIYING MA** received the B.S. degree from Xi'an Jiaotong University, Xian, China, and the M.S. degree and Ph.D. degree in electrical engineering from China Electric Power Research Institute, Beijing, China. He is currently the Deputy Director of Power System Department.

His research interests include stability analysis and control of power systems.